\documentclass[12pt]{iopart}
\usepackage[dvips]{graphicx,color}
\usepackage{times}
\begin{document}
%%%%% SYMBOL FOR NORMAL ORDERING %%%%%%%%%%%%%%%%%%%%%%
\newsavebox{\dotdot}
\savebox{\dotdot}[3mm]{\shortstack{\circle*{0.8}\\ \\ \circle*{0.8}}}
\newcommand{\nord}{\usebox{\dotdot}}
\def\pslash{{\raise0.09em\hbox{/}}\kern-0.58em \partial}

%%%%%%%%%%%%%%%%%%%%%%%%%%%%%%%%%%%%%%%%%%%%%%%%%%%%%%%%%%%
\letter{SU(N) Self-Dual Sine-Gordon Model and Competing Orders}
%%%%%%%%%%%%%%%%%%%%%%%%%%%%%%%%%%%%%%%%%%%%%%%%%%%%%%%%%%%
\author{P. Lecheminant$^1$ and K. Totsuka$^2$}
\address{$^1$ Laboratoire de Physique Th\'eorique et
Mod\'elisation, CNRS UMR 8089,
Universit\'e de Cergy-Pontoise, Site de Saint-Martin,
2 avenue Adolphe Chauvin,
95302 Cergy-Pontoise Cedex, France}
\address{$^2$ Yukawa Institute for Theoretical Physics,
Kyoto University, Kitashirakawa Oiwake-Cho, Kyoto 606-8502, Japan}
%%%%%%%%%%%%%%%%%%%%%%%%%%%%%%%%%%%%%%%%%%%%%%%%%%%%%%%%%%
\begin{abstract}
We investigate the low-energy properties of 
a generalized quantum sine-Gordon model in one dimension 
with a self-dual symmetry.  
This model describes a class of quantum phase transitions that stems from 
the competition of different orders.  
This SU($N$) self-dual sine-Gordon model
is shown to be equivalent to an SO($N$)$_2$ conformal field theory
perturbed by a current-current interaction, which is related 
to an integrable fermionic model introduced by Andrei
and Destri. In the context of spin-chain problems, 
we give several realizations of this self-dual sine-Gordon model 
and discuss the universality class of the transitions.  
\end{abstract}
%%%%%%%%%%%%%%%%%%%%%%%%%%%%%%%%%%%%%%%%%%%%%%%%%%%%%%%%%%
\pacs{75.10.Pq; 75.10.Jm;71.10.Pm}
%\noindent
\begin{center}
{\it Keywords\/} Sine-Gordon model; Self-duality; 
Competing orders; Spin chains 
\end{center}
%\newpage
%%%%%%%%%%%%%%%%%%%%%%%%%%%%%%%%%%%%%%%%%%%%%%%%%%%%%%%%%%%%

Duality symmetries have been providing much insight 
in diverse areas of physics, ranging from high-energy 
physics to condensed matter physics or statistical mechanics. 
One of the main reasons for this is that a duality
maps, in general, a theory in the strong coupling onto one in 
the weak coupling, and thus is a powerful tool for 
investigating strongly coupled regimes.  
In some lattice spin models, the duality transformation can be carried
out explicitly, mapping the partition function of one theory 
to that of another or to the same theory if the theory is  
self-dual.     
The simplest well-known example 
is the Kramers-Wannier (KW) duality transformation of the two-dimensional
Ising model, which even locates the critical point without 
calculating the partition function explicitly \cite{savit}.
In the context of the equivalent one-dimensional quantum Ising model 
in a transverse magnetic field, 
this KW duality symmetry maps the weak-field (low-temperature, in 
2D context) ordered phase onto the strong-field (high-temperature) 
paramagnetic phase and vice versa. 
Although the strong-field phase appears disordered, 
it in fact sustains a hidden order  
which is revealed by a disorder operator \cite{kadanoff}.
Since the disorder operator is usually non-local 
and dual to the standard Ising order parameter, 
the two phases, which are separated by the Ising critical point and 
characterized respectively by the order- and disorder operators, 
are in many respects rather different from each other.    
The Ising (Z$_2$) quantum phase transition that occurs in this model
can then be interpreted as a result of the competition between 
these two very different gapful orders \cite{sachdev}.  

In this letter, we shall investigate several examples
of one-dimensional competing orders whose critical
properties are described, in the continuum limit, by 
a generalization of the quantum sine-Gordon model with a manifest
self-dual symmetry.
The Hamiltonian density of the model is defined by:
\begin{equation}
\fl {\cal H}_{\rm SDSG} = 
\frac{1}{2}\left[\left(\partial_x {\vec \Phi} \right)^2
+ \left(\partial_x {\vec \Theta} \right)^2 \right]
- g \sum_{r\in \Delta^+}\left[
\nord \cos\left(\sqrt{8\pi}\, \vec{\alpha}_{r}{\cdot}{\vec\Phi}\right)
\nord
+\nord \cos\left(\sqrt{8\pi}\, \vec{\alpha}_{r}{\cdot}{\vec\Theta} \right)
\nord
\right],
\label{sunsdsgham}
\end{equation}
where the summation for $r$ is taken over the positive roots
of SU($N$) normalized to unity: ${\vec \alpha}_r^2 =1$,
and $\nord \nord$ denotes the normal ordering symbol.
The bosonic vector field ${\vec \Phi}$ is made of 
$N-1$ free boson fields $\Phi_{a}$ 
($\vec{\Phi}\equiv (\Phi_{1},\ldots,\Phi_{N-1})$)
which are defined by chiral components $\Phi_{a \rm R,L}$ as:
$\Phi_a = \Phi_{a \rm L} + \Phi_{a \rm R}$, ($a=1,\ldots,N-1$).
Similarly, each component of the dual vector-field ${\vec \Theta} = 
(\Theta_{1},\ldots,\Theta_{N-1})$ is defined by:
$\Theta_a = \Phi_{a \rm L} - \Phi_{a \rm R}$. 
The model (\ref{sunsdsgham}) is a generalization
of the usual sine-Gordon model where we have not only cosines of   
${\vec \Phi}$ but also those of the dual field  ${\vec \Theta}$.  
This field theory has been introduced in Ref. \cite{boyanovsky}
for exploring critical properties of vectorial Coulomb
gas models in the presence of both electric- and magnetic charges.
The interacting part of the model (\ref{sunsdsgham}) is marginal  
and invariant under the Gaussian duality: 
${\vec \Phi} \leftrightarrow {\vec \Theta}$ 
(i.e. the exchange of electric- and magnetic charges 
in the Coulomb gas context).  
In fact, as will be shown later, it has a hidden SO($N$) symmetry. 
Nevertheless, in what follows, 
the model (\ref{sunsdsgham}) will be referred to 
as the SU($N$) self-dual sine-Gordon (SDSG) models.  

This model is of a direct relevance to the problem of competing 
quantum orders in one dimension.  
Indeed, since $\vec{\Theta}$-field is a spatial integral of
$\partial_{t}\vec{\Phi}$, two fields $\vec{\Phi}$ and $\vec{\Theta}$ are
mutually non-local and the model (\ref{sunsdsgham}) describes,
in analogy with the above Ising duality,
the competition between two completely different orders.
In this respect, we shall give later several 
applications of the model (\ref{sunsdsgham}).  
For instance, one can anticipate that 
it describes the competition between 
a generalized charge-density wave, corresponding 
to the vertex operator of the ${\vec \Phi}$ field 
in Eq. (\ref{sunsdsgham}),
and a superconducting instability due to the perturbation
depending on the dual field.  
The exact self-duality symmetry of the model (\ref{sunsdsgham})
may suggest the existence of a non-trivial quantum criticality in the 
infrared (IR) limit that results from this competition.  
In the simplest case ($N=2$), the situation is well understood
and a Gaussian U(1) criticality emerges whatever the sign
of the coupling constant $g$ \cite{giamarchixyz,Lecheminant-G-N-02}.
This model appears in the problem of the one-dimensional
Fermi gas with backscattering and spin-non-conserving 
process as in the spin-1/2 XYZ Heisenberg chain \cite{giamarchixyz}
and also it describes critical properties of weakly-coupled
Luttinger chains \cite{Gogolin-N-T-book}.  
The low-energy property for $N > 2$ is less clear.  
The perturbative study of the model (\ref{sunsdsgham}) has been
done in Refs. \cite{boyanovsky,sierra} and a fixed point has been found
whose nature has not been fully identified in these references.  

In this letter, we shall show that for $g < 0$, 
the model (\ref{sunsdsgham}) displays a quantum critical
behavior of the level-2 SO($N$) Wess-Zumino-Novikov-Witten (WZNW) 
universality class (hereafter the level-$k$ of the Kac-Moody algebra 
will be denoted as $G_{k}$) with central charge $c=N-1$.  
In contrast, for $g > 0$, it has a fully gapped spectrum
and is related to an integrable field theory introduced 
by Andrei and Destri \cite{andrei}.

The starting point of the solution is the introduction of
$N$ right-left moving Dirac
fermions $\Psi_{\alpha R,L}, \alpha =1,\ldots,N$ with free-Hamiltonian density:
\begin{equation}
{\cal H}_0 = -i \Psi_{\alpha R}^{\dagger} \partial_x \Psi_{\alpha R}
+ i \Psi_{\alpha L}^{\dagger} \partial_x \Psi_{\alpha L} ,
\label{freediracham}
\end{equation}
where the summation over repeated indexes is assumed in the following. 
From these Dirac fermions, one can define SU($N$) ``spin''  chiral currents
through: 
\begin{equation}
J^A_{\rm R(L)} = \nord \Psi_{\alpha \rm R(L)}^{\dagger} T^A_{\alpha \beta} 
\, \Psi_{\beta \rm R(L)}\nord \; ,
\label{eqn:curre-via-Dirac}
\end{equation}
where $T^A, A=1,\ldots,N^2-1$ are the generators of 
the Lie algebra of SU($N$) in the fundamental representation
and normalized according to: $\Tr(T^A T^B)= \delta^{A B}/2$.
As well known, these currents satisfy the SU($N$)$_1$ Kac-Moody algebra 
and one can rewrite the free Hamiltonian (\ref{freediracham})
as a bilinear of currents (the so-called Sugawara form) 
\cite{Gogolin-N-T-book,CFTbook}:
\begin{equation}
\fl \quad \quad  
{\cal H}_0 =  {\cal H}_{0{\rm c}} + {\cal H}_{0{\rm s}} = 
\frac{\pi}{N}\left(\nord J_{\rm R}^2 \nord 
+ \nord J_{\rm L}^2 \nord \right)
+ \frac{2 \pi}{N+1}\left(\nord 
J^A_{\rm R} J^A_{\rm R}\nord + \nord J^A_{\rm L} J^A_{\rm L} \nord \right),
\label{freesugform}
\end{equation}
where we have introduced the U(1) ``charge'' currents: 
$J_{{\rm R(L)}} = \nord \Psi_{\alpha {\rm R(L)}}^{\dagger} 
\Psi_{\alpha {\rm R(L)}}\nord $.
At the level of the free theory ${\cal H}_{0}$, spin and charge degrees of 
freedom decouple and the free ``spin'' Hamiltonian ${\cal H}_{0{\rm s}}$
is nothing but that of the SU($N$)$_1$ WZNW conformal field theory (CFT). 
Note that the central charge of the model ${\cal H}_{0{\rm s}}$ is
$c = N -1$, i.e. the central charge of $N-1$ massless free bosons which
describes the $g\rightarrow 0$ limit of the SDSG model (\ref{sunsdsgham}).

Now let us add a perturbation ${\cal V}$ to the ``spin'' (or SU($N$))
Hamiltonian ${\cal H}_{\rm 0s}$ so that the ``spin'' part 
${\cal H}_{\rm 0s}+{\cal V}\equiv {\cal H}_{N}$ coincide with 
the sine-Gordon model (\ref{sunsdsgham}).  
Obviously, it should be marginal (i.e. four fermion interaction) and 
invariant under both chiral (R$\leftrightarrow$L) symmetry 
and the Gaussian duality $\vec{\Phi}\leftrightarrow \vec{\Theta}$. 
This self-duality symmetry considerably restricts the form of the possible
four fermion interactions.  
To see this, let us introduce $N$ chiral bosonic fields
$\varphi_{\alpha \rm R,L}$ using the Abelian bosonization of 
Dirac fermions \cite{Gogolin-N-T-book}: 
\begin{eqnarray}
\Psi_{\alpha {\rm R}} &=& \frac{\kappa_{\alpha}}{\sqrt{2\pi}}
\nord \exp\left(i \sqrt{4\pi} \; \varphi_{\alpha {\rm R}}\right) \nord 
\label{bosonization-1}
\\
\Psi_{\alpha {\rm L}} &=& \frac{\kappa_{\alpha}}{\sqrt{2\pi}}
\nord \exp\left( - i \sqrt{4\pi} \; \varphi_{\alpha {\rm L}}\right) \nord ,
\label{bosonization-2}
\end{eqnarray}
where the bosonic fields satisfy the commutation 
relation $[\varphi_{\alpha {\rm R}}, \varphi_{\beta {\rm L}} ]
= i \delta_{\alpha \beta}/4$. The anticommutation between fermions 
with different indexes
is realized through the presence of 
Klein factors (here Majorana fermions) 
$\kappa_{\alpha}$ with the following anticommutation 
rule: $\{\kappa_{\alpha}, \kappa_{\beta} \} = 
2 \delta_{\alpha \beta}$.
The Gaussian duality 
symmetry: $\varphi_{\alpha} (\equiv \varphi_{\alpha {\rm L}} 
+  \varphi_{\alpha {\rm R}}) \leftrightarrow
\vartheta_{\alpha} (\equiv \varphi_{\alpha {\rm L}} 
-  \varphi_{\alpha {\rm R}})$ thus amounts to 
the particle-hole (P-H) transformation {\em only} in the right-moving 
(R) sector of the Dirac theory: 
$\Psi_{\alpha {\rm R}} \rightarrow \Psi_{\alpha {\rm R}}^{\dagger}$, 
$\Psi_{\alpha {\rm L}} \rightarrow \Psi_{\alpha {\rm L}}$.
As is well known, the SU($N$) generators $T^A$ can be classified 
into three categories: 
\begin{itemize}
\item {\rm Antisymmetric i.e. SO($N$) part:}
\begin{equation}
\left(T^{\rm{SO(N)}}_{ij}\right)_{\alpha\beta}
=-\frac{i}{2}(\delta_{i\alpha}\delta_{j\beta}
-\delta_{i\beta}\delta_{j\alpha})
\qquad (1\leq i < j \leq N)
\label{eqn:SUN-gen-1}
\end{equation}
\item {\rm Symmetric part:}
\begin{equation}
\left(T^{\rm{S}}_{ij}\right)_{\alpha\beta}
=\frac{1}{2}(\delta_{i\alpha}\delta_{j\beta}
+ \delta_{i\beta}\delta_{j\alpha})
\qquad (1\leq i < j \leq N)
\label{eqn:SUN-gen-2}
\end{equation}
\item {\rm Cartan generators (Diagonal):}
\begin{equation}
\fl \left(T^{\rm{D}}_{m}\right)_{\alpha\beta}
=\frac{1}{\sqrt{2m(m+1)}}
\left(
\sum_{k=1}^{m}\delta_{\alpha k}\delta_{\beta k}
-m\, \delta_{\alpha,m+1}\delta_{\beta,m+1}
\right),  (m=1,\ldots,N-1) .
\label{eqn:SUN-gen-3}
\end{equation}
\end{itemize}

Since all generators belonging to
the SO($N$) subset are antisymmetric, we deduce that the corresponding right 
currents behave under the Gaussian 
duality as:
\begin{eqnarray}
J^{\rm{SO(N)}}_{{\rm R},ij} &=&
\nord \Psi_{\alpha {\rm R}}^{\dagger}
\left(T^{\rm{SO(N)}}_{ij}\right)_{\alpha\beta}
\Psi_{\beta {\rm R}} \nord
\nonumber \\
&\stackrel{\rm{P-H}}{\longrightarrow} &
- \nord \Psi_{\beta {\rm R}}^{\dagger}
\left(T^{\rm{SO(N)}}_{ij}\right)_{\alpha\beta}
\Psi_{\alpha {\rm R}} \nord = J^{\rm{SO(N)}}_{{\rm R},ij} ,
\end{eqnarray}
whereas the remaining $(N+2)(N-1)/2$ SU($N$) generators are all symmetric
or diagonal
and the corresponding right currents change sign 
under the Gaussian duality:
$J^{\rm S,D}_{\rm R} \stackrel{\rm{P-H}}{\longrightarrow}  
-J^{\rm S,D}_{\rm R}$. 
In contrast, the Gaussian duality does nothing for $J_{\rm L}^{A}$. 
This argument suggests that a possible model equivalent to 
the SDSG model (\ref{sunsdsgham}) might be
\begin{eqnarray}
{\cal H}_N &=&
\frac{2 \pi}{N+1}\sum_{A\in {\rm SU(N)}}\left(\nord 
J^A_{\rm R} J^A_{\rm R}\nord + \nord J^A_{\rm L} J^A_{\rm L} 
\nord\right) + \lambda \, \sum_{i<j}^{N}
J_{{\rm R},ij}^{\rm{SO(N)}} J_{{\rm L},ij}^{\rm{SO(N)}}  \nonumber \\
&=&  
\frac{4\pi}{N}\sum_{i<j}^{N}
\left[\nord ( J_{{\rm R},ij}^{\rm{SO(N)}})^{2}\nord +
\nord ( J_{{\rm L},ij}^{\rm{SO(N)}})^{2}\nord \right]
+ \lambda \, \sum_{i<j}^{N}
J_{{\rm R},ij}^{\rm{SO(N)}}
 J_{{\rm L},ij}^{\rm{SO(N)}} , 
\label{hamguess}
\end{eqnarray}
for an appropriately chosen coupling constant $\lambda$.  
In fact, by using bosonization rules (\ref{bosonization-1})-%
(\ref{bosonization-2}), we can derive the SDSG model (\ref{sunsdsgham}) 
from (\ref{hamguess}).   
Plugging Eqs. (\ref{eqn:curre-via-Dirac}), (\ref{bosonization-1}) 
and (\ref{bosonization-2}) into (\ref{hamguess}), one obtains 
\begin{eqnarray}
{\cal H}_N &=& \frac{1}{N} \sum_{i<j}
\left[ \left(\partial_{x}\varphi_{i {\rm R}}
-\partial_{x}\varphi_{j {\rm R}}\right)^{2}
+ \left(\partial_{x}\varphi_{i {\rm L}}
-\partial_{x}\varphi_{j {\rm L}}\right)^{2} \right]
\nonumber \\
& & - \frac{\lambda}{8 \pi^2} \sum_{i<j}
\left\{
\nord \cos (\sqrt{4\pi}(\varphi_{i}-\varphi_{j})) \nord
+ \nord \cos (\sqrt{4\pi}(\vartheta_{i}-\vartheta_{j})) \nord
\right\} . 
\label{hamguessbose}
\end{eqnarray}
If we introduce a charge bosonic field
$\Phi_{c{\rm R,L}}$ and the SU($N$) bosonic fields $\Phi_{a {\rm R,L}}$
($a = 1, \ldots, N-1$) as \cite{assaraf}: 
\begin{eqnarray}
\Phi_{c {\rm R(L)}} &=& \frac{1}{\sqrt{N}}\left(
\varphi_1 + \ldots + \varphi_N \right)_{\rm R(L)}
\nonumber \\ 
\Phi_{a {\rm R(L)}} &=& \frac{1}{\sqrt{a(a+1)}}\left(
\varphi_1 + \ldots + \varphi_a - a  \varphi_{a+1}\right)_{\rm R(L)} \; ,
\label{SUNbasis}
\end{eqnarray}
the non-interacting part of 
Eq. (\ref{hamguessbose}) takes the standard form 
of a kinetic term for free bosons and 
the Hamiltonian (\ref{hamguessbose}) reads 
\begin{equation}
\fl {\cal H}_{N} =
\frac{1}{2}\left[\left(\partial_x {\vec \Phi} \right)^2
+ \left(\partial_x {\vec \Theta} \right)^2 \right]
-  \frac{\lambda}{8 \pi^2} \sum_{r\in \Delta^+}\left\{
\nord \cos\left(\sqrt{8\pi}\, \vec{\alpha}_{r}{\cdot}{\vec\Phi}\right)
\nord
+\nord \cos\left(\sqrt{8\pi}\, \vec{\alpha}_{r}{\cdot}{\vec\Theta} \right)
\nord
\right\} \; .
\label{sunsdsghambis}
\end{equation}
Thus we have shown that the model (\ref{hamguess}) is indeed 
equivalent to the SDSG model (\ref{sunsdsgham}) if we identify 
$\lambda = 8 \pi^2 g$. 

To deduce the physical properties
of the model (\ref{hamguess}), 
it is more enlightening to introduce $2N$ Majorana
fermions $\xi_{\rm R,L}^i$ and $\chi_{\rm R,L}^i$ 
($i =1, \ldots, N$) from the Dirac ones:
$\Psi_{i \rm R,L} = (\xi_{\rm R,L}^i + i\, \chi_{\rm R,L}^i)/\sqrt{2}$.
The $J_{{\rm R(L)},ij}^{\rm{SO(N)}}$, being bilinears of Dirac fermions,
can be expressed in terms of the Majorana fermions:
\begin{equation}
J_{{\rm R(L)},ij}^{\rm{SO(N)}} = 
-\frac{i}{2}\left(
\xi_{\rm R(L)}^i\xi_{\rm R(L)}^j+
\chi_{\rm R(L)}^i \chi_{\rm R(L)}^j
\right) = \frac{1}{2} {\cal J}_{{\rm R(L)},ij}^{\rm{SO(N)}} , 
\label{so2ncurrents}
\end{equation}
where ${\cal J}_{{\rm R(L)},ij}^{\rm{SO(N)}}$, being the sum
of two SO($N$)$_1$ currents, is an SO($N$)$_2$ current. 
Therefore, we deduce that 
the SDSG model (\ref{sunsdsgham}) is equivalent to 
the level-2 SO($N$) WZNW model
perturbed by a marginal current-current interaction:   
\begin{equation}
\fl \qquad \qquad
{\cal H}_{\rm SDSG}
= \frac{\pi}{N}
\sum_{i<j}
\left[\nord ({\cal J}_{{\rm R},ij}^{\rm{SO(N)}})^{2}\nord +
\nord ({\cal J}_{{\rm L},ij}^{\rm{SO(N)}})^{2}\nord \right]
+ 2\pi^{2} g \, \sum_{i<j}
{\cal J}_{{\rm R},ij}^{\rm{SO(N)}}
{\cal J}_{{\rm L},ij}^{\rm{SO(N)}} .
\label{so2ncurentcurrent}
\end{equation}
This equation is one of the main results of this letter.  

Using this equivalence, one can extract the IR
properties of the SDSG model.  
The one-loop renormalization-group (RG) equation 
of the model (\ref{so2ncurentcurrent}) is: 
${\dot g} =  (N - 2) \pi  g^2$, where ${\dot g}= \partial g/\partial l$
($l$ being the RG parameter).
For $g < 0$, the interaction is marginally irrelevant
so that in the far IR limit, the model flows towards
the SO($N$)$_2$ WZNW fixed point with a central charge
$c=N-1$. This CFT has the same
central charge as the SU($N$)$_1$ WZNW model and in fact there exists
a conformal embedding between them \cite{CFTbook}:
SU$(N)_{1}\,{\supset}$ SO$(N)_{2}$.  
In contrast, when $g > 0$, the interaction is marginally
relevant and flows toward strong coupling.  
From the structure of the current-current interaction, it is 
naturally expected
that a mass gap opens dynamically i.e. the SDSG model
is a massive field theory for $g > 0$ and $N > 2$.  
In fact, this can be explicitly shown by observing
that model (\ref{so2ncurentcurrent}) is related
to an integrable field theory introduced
by Andrei and Destri \cite{andrei} (see also
Ref. \cite{Reshetikhin}) with the following Hamiltonian:
\begin{equation}
{\cal H}_{\rm AD} = -\frac{i}{2}\left(
{\bar \psi}_{1,i} \; \gamma^{1}\partial_{x}  \psi_{1,i}
+ {\bar \psi}_{2,i} \; \gamma^{1}\partial_{x} \psi_{2,i}
\right) 
- g_{\rm AD} ( \rho^2 
+ \tilde{\rho}^2 + \sigma^2 + \tilde{\sigma}^2) ,
\label{ADlagrangian}
\end{equation}
where $\psi_{1,i}$ (respectively $\psi_{2,i}$)
is a two-component spinor formed by $\xi^i_{\rm R,L}$
(respectively $\chi^i_{\rm R,L}$) and $\gamma^0 = \sigma_2$,
$\gamma^1 = i \sigma_1$, and $\gamma_5 = \sigma_3$
($\sigma_i$ being the Pauli matrices).  
The O($N$)-invariant order parameters $\rho$, $\sigma$, 
$\tilde{\rho}$ and $\tilde{\sigma}$ are defined as:
\begin{eqnarray}
\rho &\equiv& \frac{1}{2}\left(
\bar{\psi}_{1,i}\psi_{1,i}+\bar{\psi}_{2,i}\psi_{2,i}
\right)
=-i\left(
\xi^{i}_{\rm R}\xi^{i}_{\rm L}
+\chi^{i}_{\rm R}\chi^{i}_{\rm L}
\right) 
\label{orderparameter1}
\\
\sigma &\equiv &
-\bar{\psi}_{1,i}\gamma^{5}\psi_{2,i}
=-i\left(
\xi^{i}_{\rm R}\chi^{i}_{\rm L}
-\chi^{i}_{\rm R}\xi^{i}_{\rm L}
\right)
\label{orderparameter2}
\\
\tilde{\rho} &\equiv& \frac{1}{2}\left(
\bar{\psi}_{1,i}\psi_{1,i}-\bar{\psi}_{2,i}\psi_{2,i}
\right)
=-i\left(
\xi^{i}_{\rm R}\xi^{i}_{\rm L}
-\chi^{i}_{\rm R}\chi^{i}_{\rm L}
\right)  
\label{orderparameter3}
\\
\tilde{\sigma} &\equiv&
\bar{\psi}_{1,i}\psi_{2,i}
= -i\left(
\xi^{i}_{\rm R}\chi^{i}_{\rm L}
+ \chi^{i}_{\rm R}\xi^{i}_{\rm L}
\right) . 
\label{orderparameter4}
\end{eqnarray}
Although ${\cal H}_{\rm AD}$ looks complicated, after bosonizing, 
the Hamiltonian (\ref{ADlagrangian}) 
separates into two commuting pieces, a free Hamiltonian 
${\cal H}_{0\rm c}$ for the massless bosonic field $\Phi_{\rm c}$ 
(Eq. (\ref{SUNbasis})) and the SO($N$)$_2$ current-current model 
(\ref{so2ncurentcurrent}) with $g= g_{\rm AD}/\pi^2$:
${\cal H}_{\rm AD} = {\cal H}_{0\rm c} + {\cal H}_{\rm SDSG}$.
Since the model (\ref{ADlagrangian}) is exactly solvable 
by means of the Bethe ansatz \cite{andrei,Reshetikhin}, 
we can extract the physical properties of ${\cal H}_{\rm SDSG}$ 
from the solution.  
The nature of the ground states may be simply understood in terms of
the order parameters $\rho$, $\sigma$, $\tilde{\rho}$ and
$\tilde{\sigma}$,    
which form two independent SO(2) doublets
$(\rho,\sigma)$ and $(\tilde{\rho},\tilde{\sigma})$
\cite{andrei,Lecheminant-T-05}.   
These two doublets are mapped onto each other by the KW duality  
for the Majorana fermions $\chi^i$: 
$\chi^i_{\rm R} \rightarrow - \chi^i_{\rm R}$ and 
$\chi^i_{\rm L} \rightarrow \chi^i_{\rm L}$.  
From the form of the interacting part of the model (\ref{ADlagrangian}), 
we readily see that 
it is invariant under the interchange of the two doublets.   
On the basis of large-$N$ semiclassical argument, the authors 
of Ref. \cite{andrei} found 
that when $g_{\rm AD} <0$, this interchange symmetry is broken  
spontaneously in the ground state and that there are two different  
ground states where only one of the two doublets has a finite modulus;   
correspondingly massive kink excitations appear in the spectrum 
to connect the above ground states.   

%%%%%%%%%%%%%%%%%%%%%%%%%%%%%%%%%%%%%%%%%%%%%%%%%%%%%%%%%%%%%%%%
Let us now consider some physical applications 
of the SDSG model (\ref{sunsdsgham}) 
in the context of competing orders in one dimension.   
As the first example, 
we take a spin-1 bilinear-biquadratic Heisenberg chain 
\cite{Affleck-K-L-T-88} with nearest ($J_1$) and next-nearest ($J_2$) 
interactions:
\begin{eqnarray}
H_{\rm BB} &=&
\sum_n\sum_{a=1}^2 J_{\rm BB}^{(a)} \left[ {\bf S}_n \cdot {\bf S}_{n+a} 
+ \left({\bf S}_n \cdot {\bf S}_{n+a}\right)^2 \right]
+ \delta_{\rm BB} \sum_n {\bf S}_n \cdot {\bf S}_{n+1} ,
\label{spin1ham}
\end{eqnarray}
with ${\bf S}_n$ being a spin-1 operator at site $n$. 
The model with $\delta_{\rm BB}=0$ is SU(3)-symmetric and, in particular, 
for $J_{\rm BB}^{(2)}=0$ it reduces to an integrable model 
\cite{uimin,sutherland} which displays a quantum critical behavior
of the SU(3)$_1$ WZNW universality class \cite{affleck,itoi}.  
The effect of the remaining interactions 
can be investigated in the vicinity of the SU(3) symmetric point 
($J_{\rm BB}^{(2)}=\delta_{\rm BB}=0$) using the low-energy approach of 
Ref. \cite{itoi}. In fact, Itoi and Kato \cite{itoi} considered 
a more general problem of an SU($N$)$_1$ WZNW model perturbed 
by the {\em most general} SO($N$)-symmetric marginal perturbation:
\begin{eqnarray}
\fl
{\cal H}_{\rm IK} = 
\frac{2 \pi}{N+1}\left(\nord J^A_{\rm R} J^A_{\rm R}\nord 
+ \nord J^A_{\rm L} J^A_{\rm L} \nord\right)
+ \lambda_1 J^A_{\rm R} J^A_{\rm L} 
+ 2 \lambda_2
\left(T^A_{\alpha \beta} T^B_{\alpha \beta} \right)
J^B_{\rm R} J^A_{\rm L} ,
\label{itoimodel}
\end{eqnarray}
which for $N=3$ should describe the low-energy physics of 
the SO(3) model (\ref{spin1ham}) around the SU(3) symmetric point.  

\begin{figure}[h]
\begin{center}
\includegraphics[scale=0.7]{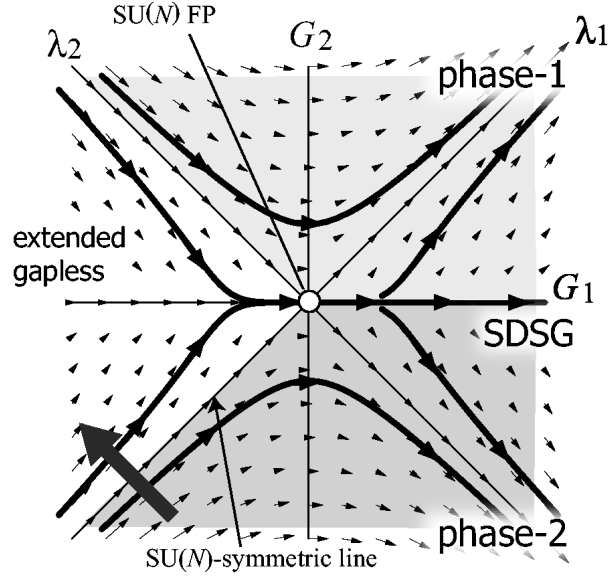}
\end{center}
\caption{One-loop RG flow for the Itoi-Kato model (\ref{itoimodel}).   
In the sine-Gordon language, the $\lambda_{1}$- and the $\lambda_{2}$ axis 
respectively correspond to the pure 
$\cos(\sqrt{8\pi}\vec{\alpha}_r{\cdot}\vec{\Phi})$ model 
and the pure $\cos(\sqrt{8\pi}\vec{\alpha}_r{\cdot}\vec{\Theta})$ one 
(see Eq. (\ref{sunsdsgham})).  Since they are related to each other by 
the Gaussian duality, {\em phase-1} and {\em phase-2} cannot be connected 
by any local symmetry.  
These competing gapful phases are separated by 
the line of self-dual sine-Gordon model (SDSG).  
The thick arrow schematically shows the path traced by 
$H_{\rm BB}(J_{\rm BB}^{(2)}=0)$ 
($H_{\rm so}(J_{\rm so}^{(2)}=0)$ for $N=4$) as 
$\delta_{\rm BB}$ ($\delta_{\rm so}$ for $N=4$) is changed from positive 
(Haldane- or staggered dimerization phase)
to negative (gapless SU($N$)$_{1}$ WZNW phase).\label{fig:RG-flow}}
\end{figure}

Using the decomposition 
(\ref{eqn:SUN-gen-1},\ref{eqn:SUN-gen-2},\ref{eqn:SUN-gen-3}) 
of the SU($N$) generators $T^A$, the model (\ref{itoimodel})
can be expressed in the following compact form:
\begin{equation}
\fl
{\cal H}_{\rm IK} =
\frac{2 \pi}{N+1}\left(
\nord J^A_{\rm R} J^A_{\rm R}\nord 
+ \nord J^A_{\rm L} J^A_{\rm L}\nord  \right)
+ \left(\lambda_1 -\lambda_2 \right) 
\sum_{A\in\rm{SO(N)}}J^A_{\rm R} J^A_{\rm L}
+ \left(\lambda_1 + \lambda_2 \right)
\sum_{A\in\rm{S,D}} J^A_{\rm R} J^A_{\rm L} .
\label{itoimodelbis}
\end{equation}
It is straightforward to calculate the one-loop RG  
equations \cite{itoi} for the model (\ref{itoimodelbis}) and we obtain 
\begin{equation}
\dot{G}_{1} = \frac{N-2}{8\pi}G_{1}^{2}+\frac{N+2}{8\pi}G_{2}^{2} 
\; , \quad 
\dot{G}_{2} = \frac{N}{4\pi}G_{1}G_{2} \; ,
\label{eqn:RGE}
\end{equation}
where we have introduced a new set of couplings as
$G_{1}\equiv \lambda_{1}-\lambda_{2}$ and 
$G_{2}\equiv \lambda_{1}+\lambda_{2}$.   The RG-flow 
is shown in Figure \ref{fig:RG-flow}.  We have two gapful phases 
({\em phase-1} and {\em phase-2}) together with one extended gapless 
phase which belongs to the SU($N$)$_{1}$ WZNW universality class.   

In the special case of $N=3$, the Hamiltonian (\ref{itoimodelbis}) 
describes the competition between two gapful orders 
of $H_{\rm BB}$: a trimerization (period-3) phase ({\em phase-1} 
in Figure \ref{fig:RG-flow}) 
stabilized when $\lambda_2 = 0$ and $\lambda_1 > 0$, 
where three adjacent spins form local SU(3) singlets, 
and the non-degenerate Haldane state ({\em phase-2}) 
when $\lambda_1 = 0$ and $\lambda_2 < 0$ \cite{itoi,comment}.
The trimerized phase is expected to occur in $H_{\rm BB}$ 
when $\delta_{\rm BB}=0$ and for a sufficiently strong value of 
$J_{\rm BB}^{(2)}$, 
whereas the Haldane phase appears 
when $J_{\rm BB}^{(2)} =0$ and $\delta_{\rm BB} > 0$ \cite{Affleck-K-L-T-88}.
The one-loop RG flow is presented in Figure \ref{fig:RG-flow} and
we see that the phase transition between these two gapful phases occurs
along the line $\lambda_1 = - \lambda_2 > 0$  
shown as `SDSG'.   
From Eq. (\ref{itoimodelbis}), we find that the effective field theory  
which describes the transition is given by the SO($N$) current-current 
model (\ref{so2ncurentcurrent}) with $N=3$
and $g = \lambda_1/4\pi^2 (=-\lambda_2/4\pi^2) > 0$.
We thus have found an example of the SDSG model 
(\ref{sunsdsgham}) with $N=3$ 
which describes the competition between the Haldane and trimerized orders
and corresponds to a first-order transition (since the gap 
opens for $g > 0$).
%%%%%%%%%%%%% FIG %%%%%%%%%%%%%%%%%%%%%%%%%%%%%%%%%%%%%%%%%%%%%%
%%%%%%%%%%%%%%%%%%%%%%%%%%%%%%%%%%%%%%%%%%%%%%%%%%%%%%%%%%%%%%%%

%%%%%%%%%%%% SU(4) %%%%%%%%%%%%%%%%%%%%%%%%%%%%%%%%%%%%%%%%%%%%%%%
In the second example, we consider 
an SU(2)$\times$SU(2)-symmetric spin-orbital chain 
with nearest ($J_{\rm so}^{(1)}$) and next-nearest ($J_{\rm so}^{(2)}$) 
interactions:
\begin{equation}
\fl
H_{\rm so} = 
\sum_n \sum_{a=1}^{2} J^{(a)}_{\rm so}
\left( 2\, {\bf S}_{n} {\cdot} {\bf S}_{n+a}
+ \frac{1}{2} \right) \left( 2\, {\bf T}_{n} {\cdot} {\bf T}_{n+a}
+ \frac{1}{2} \right) + 
\delta_{\rm so} \sum_n \left({\bf S}_{n} {\cdot} {\bf S}_{n+1}
+ {\bf T}_{n} {\cdot} {\bf T}_{n+1}\right) ,
\label{spinorbitalmodel}
\end{equation}
where ${\bf S}_{n}$ and ${\bf T}_{n}$ denote spin-1/2
operators representing respectively 
the spin- and the two-fold degenerate orbital 
degrees of freedom \cite{pati,yamashita} on the $n$-th site. 
For $J^{(2)}_{\rm so}=\delta_{\rm so}=0$, 
the model coincides with an SU(4) generalization of 
the spin-1/2 Heisenberg chain and is exactly solvable by 
Bethe ansatz \cite{sutherland};   
the model is gapless with three massless bosonic modes and the field theory 
describing this quantum criticality is the 
SU(4)$_1$ WZNW model \cite{affleck,Itoi-Q-A-00} 
or, equivalently, the SO(6)$_1$ WZNW theory in terms of two triplets of 
Majorana fermions $\xi^i_{\rm R,L},\chi^i_{\rm R,L}, i=1,\ldots,3$ 
\cite{azaria,Lecheminant-T-05}.  
Using the Majorana basis, one can derive the low-energy effective
Hamiltonian 
of the model (\ref{spinorbitalmodel}) in the vicinity of the SU(4) point 
($J^{(2)}_{\rm so}=\delta_{\rm so}=0$). 
We find, using the results of Refs. \cite{Lecheminant-T-05,azaria},
the following effective Hamiltonian density:
\begin{eqnarray}
{\cal H}_{\rm so} &=&
- \frac{ iv}{2}
\left( \xi_{\rm R}^i \partial_x \xi_{\rm R}^i
- \xi_{\rm L}^i \partial_x \xi_{\rm L}^i
+ \chi_{\rm R}^i \partial_x \chi_{\rm R}^i
- \chi_{\rm L}^i \partial_x \chi_{\rm L}^i \right)
\nonumber \\ 
&+& \left(g_1 + g_2 \right) \left( \xi_{\rm R}^i \xi_{\rm L}^i 
+ \chi_{\rm R}^i \chi_{\rm L}^i \right)^2
+ \left(g_1 - g_2 \right)  \left( \xi_{\rm R}^i \xi_{\rm L}^i 
- \chi_{\rm R}^i \chi_{\rm L}^i \right)^2 .
\label{spinorbitalhamcont}
\end{eqnarray}
Since the interacting part of Eq. (\ref{spinorbitalhamcont})  
can be written as: 
${\cal H}_{\rm so}^{\rm int}
= -(g_{1}+g_{2})\, \rho^{2}-(g_{1}-g_{2})\tilde{\rho}^{2}$, 
we may expect that the model describes the competition between two 
different fully gapped orders which are characterized, 
in this continuum limit, 
by the order parameters $\rho$ and ${\tilde \rho}$ 
of Eqs. (\ref{orderparameter1},\ref{orderparameter3}).  
As has been shown in Ref. \cite{Lecheminant-T-05},  
they correspond respectively to an SU(4) quadrumerization (Q) phase 
(i.e. period 4), which is characterized by local SU(4) singlets, and to 
a period-2 staggered dimerization (SD) phase which is formed by 
alternating spin and orbital singlets \cite{nersesyan}.  
In terms of the lattice coupling constants, the former phase 
emerges when $\delta_{\rm so} =0$ and for a sufficiently 
large value of $J_{\rm so}^{(2)}$ \cite{lauchli} 
whereas the SD phase is stabilized when $J_{\rm so}^{(2)} = 0$ 
and $\delta_{\rm so}, J_{\rm so}^{(1)} > 0$ \cite{pati}.  
The competition between these two orders can be investigated 
by observing that 
the SU(2)$\times$SU(2)-symmetric model ${\cal H}_{\rm so}$ 
(\ref{spinorbitalmodel}) can be recasted into the form 
of ${\cal H}_{\rm IK}$ (\ref{itoimodelbis}) with $N=4$    
since SU(2)$\times$SU(2)$\sim$SO(4). 
In fact, an explicit calculation shows 
\begin{equation}
{\cal H}_{\rm so}^{\rm int} = 
8g_{1}\sum_{A\in \rm SO(4)}J_{\rm R}^{A}J_{\rm L}^{A} 
+ 8 g_{2}\sum_{A\in\rm S,D}J_{\rm R}^{A}J_{\rm L}^{A} \; .
\end{equation}
With the identification $G_{1,2}=8 g_{1,2}$ and $N=4$, the RG equations 
(\ref{eqn:RGE}) again describe the model ${\cal H}_{\rm so}$.   
From the Figure \ref{fig:RG-flow},
we observe that the quantum phase transition 
between the two competing phases (phase-1 for `Q' and phase-2 for `SD') 
occurs at $g_2 = 0$ and $g_1 > 0$ and is thus described by:
\begin{equation}
{\cal H}_{\rm so}(g_{2}=0) 
= {\cal H}_{\rm IK}(\lambda_{1}=-\lambda_{2}=4g_{1})
= {\cal H}_{\rm SDSG}^{N=4}(g=g_{1}/\pi^{2})  \; .
\end{equation}
From the equivalence between ${\cal H}_{\rm SDSG}$ and 
the massive sector of ${\cal H}_{\rm AD}$, we conclude that 
the Q$\leftrightarrow$SD transition described by 
${\cal H}_{\rm so}(g_{2}=0)$ is of first order.   

The last example is provided by the generalized two-leg 
spin ladders with four-spin exchange interactions
studied recently in Ref. \cite{Lecheminant-T-05}.
In particular, it has been shown that, close to the 
SU(4) symmetric point of Eq. (\ref{spinorbitalmodel})
with $J_{\rm so}^{(2)} = \delta_{\rm so} =0$, four competing orders emerge.  
In addition to the Q ($\rho$) and SD ($\tilde{\rho}$) 
orders of the previous example, 
a scalar-chirality order \cite{lauchliST} and 
a rung-quadrumerization order appear.   
These two additional phases are characterized, within
the low-energy approach of Ref. \cite{Lecheminant-T-05},
respectively by the order parameters $\sigma$ and 
${\tilde \sigma}$ of Eqs. (\ref{orderparameter2},\ref{orderparameter4}).
As is seen from the one-loop RG analysis of Ref. \cite{Lecheminant-T-05},
the competition between these four orders is governed by
the low-energy effective Hamiltonian:
\begin{equation}
\fl \qquad 
{\cal H}_{\rm eff} =
- \frac{ iv}{2}
\left( \xi_{R}^i \partial_x \xi_{R}^i
- \xi_{L}^i \partial_x \xi_{L}^i
+ \chi_{R}^i \partial_x \chi_{R}^i
- \chi_{L}^i \partial_x \chi_{L}^i \right)
- \lambda \left(
\rho^2 + {\tilde \rho}^2
+ \sigma^2 + {\tilde \sigma}^2
\right),
\label{fourcompetingorders}
\end{equation}
with $\lambda > 0$. We thus observe that the Andrei-Destri
model (\ref{ADlagrangian}) with $N=3$ and $g_{\rm AD} = \lambda$ 
accounts for 
the competition between the four different orders of the problem.
Since the latter model is equivalent to the SDSG model (\ref{sunsdsgham}) 
up to a free massless bosonic field, we easily see that 
the resulting phase transition is of a U(1) Gaussian type when
$\lambda > 0$.
%%%%%%%%%%%%%%%%%%%%%%%

%%%%%%%%%%%%%%%%%%%%%%%

We hope that other applications of the SDSG model will be
reported in the near future.
\ack
\qquad
We would like to thank H. Saleur for illuminating discussions
and N. Andrei, P. Azaria, E. Boulat, and A. A. Nersesyan  
for their interest in this work.  We are also grateful to 
A.~L\"{a}uchli for sharing his unpublished results and 
for useful comments on the manuscript.   
The author (K.T.) is grateful to the members of LPTM at 
Cergy-Pontoise university where this work has been carried out.  
He is supported in part by the Grant-in-Aid No. 18540372 from 
MEXT of Japan.  
%%%%%%%%%%%%%%%%%%%%%%%%%%%%%%%%%%%%%%%%%%%%%%%%%%%%%%%%%%h


\begin{thebibliography}{10}
\bibitem{comment}
The ground-state degeneracy is not discussed in the original
treatment \cite{itoi}.
We derived an effective Hamiltonian starting from the SU(3) Hubbard
chain and counted the number of inequivalent (semiclassical) ground
states.
\end{thebibliography}
\end{document}